\documentclass[a4paper]{spie}  

\usepackage{amsmath,amsfonts,amssymb,aas_macros, url}
\usepackage{graphicx}
\usepackage[colorlinks=true, allcolors=blue]{hyperref}

\title{The CUbesat Solar Polarimeter (CUSP) mission overview}

\author[a]{Sergio~Fabiani}
\author[a]{Ettore~Del~Monte}
\author[b]{Ilaria~Baffo}
\author[c]{Sergio~Bonomo}
\author[g]{Daniele~Brienza}
\author[e]{Riccardo~Campana}
\author[h]{Mauro~Centrone}
\author[d]{Gessica~Contini}
\author[a]{Enrico~Costa}
\author[c]{Giovanni~Cucinella}
\author[f]{Andrea~Curatolo}
\author[a]{Nicolas~De~Angelis}
\author[e]{Giovanni~De~Cesare}
\author[d]{Andrea~Del~Re}
\author[a]{Sergio~Di~Cosimo}
\author[c]{Simone~Di~Filippo}
\author[a]{Alessandro~Di~Marco}
\author[a]{Giuseppe~Di~Persio}
\author[g]{Immacolata~Donnarumma}
\author[b]{Pierluigi~Fanelli}
\author[d]{Paolo~Leonetti}
\author[f]{Alfredo~Locarini}
\author[a]{Pasqualino~Loffredo}
\author[a,i]{Giovanni~Lombardi}
\author[l]{Gabriele~Minervini}
\author[f]{Dario~Modenini}
\author[a]{Fabio~Muleri}
\author[g]{Silvia~Natalucci}
\author[c]{Andrea~Negri}
\author[c]{Massimo~Perelli}
\author[a]{Monia~Rossi}
\author[a]{Alda~Rubini}
\author[a]{Emanuele~Scalise}
\author[a]{Paolo~Soffitta}
\author[g]{Andrea~Terracciano}
\author[f]{Paolo~Tortora}
\author[g]{Emauele~Zaccagnino}
\author[d]{Alessandro~Zambardi}

\affil[a]{INAF-IAPS, via del Fosso del Cavaliere 100, 00133 Rome, Italy}
\affil[b]{DEIM, Universit\'a degli studi della Tuscia, Largo dell'Universit\'a, 01100 Viterbo, Italy}
\affil[c]{IMT s.r.l., via Carlo Bartolomeo Piazza 30, 00161 Rome, Italy}
\affil[d]{SCAI Connect s.r.l.,  Via Francesco Gentile 135, 00173 Rome, Italy}
\affil[e]{INAF-OAS Bologna, via Piero Gobetti 93/3, 40129 Bologna, Italy}
\affil[f]{Alma Mater Studiorum Universit\'a di Bologna - Department of Industrial Engineering and Interdepartmental Center for Industrial Aerospace Research, Via Fontanelle 40, 47121 Forl\'i, Italy}
\affil[g]{ASI, Via del Politecnico snc 00133 - Roma, Italy}
\affil[h]{INAF-OAR, Via Frascati 33, 00040, Monte Porzio Catone, Italy}
\affil[i]{ Dipartimento di Ingegneria dell’Impresa ''Mario Lucenti", Università degli Studi di Roma ``Tor Vergata”, Via Cracovia 50, 00133 Roma, Italy}
\affil[l]{ INAF Headquarters, Viale del Parco Mellini 84, 00136, Roma, Italy}

\authorinfo{Further author information: (Send correspondence to Sergio Fabiani)\\Sergio Fabiani: E-mail: sergio.fabiani@inaf.it, Telephone: +39 06 4993 4450}

\begin{document}
\maketitle

\begin{abstract}
The CUbesat Solar Polarimeter (CUSP) project is a future CubeSat mission orbiting the Earth aimed to measure the linear polarization of solar flares in the hard X-ray band, by means of a Compton scattering polarimeter. CUSP will allow us to study the magnetic reconnection and particle acceleration in the flaring magnetic structures of our star. The project is in the framework of the Italian Space Agency Alcor Program, which aims to develop new CubeSat missions. 
CUSP is approved for a Phase B study that will last for 12 months,  starting in mid-2024.
We report on the current status of the CUSP mission project as the outcome of the Phase A.
\end{abstract}


\section{INTRODUCTION}
\label{sec:intro}  
Solar activity can have a significant impact on human technology, both on ground and in space. Violent phenomena such as Solar Flares (SFs) release a huge amount of energy in the solar corona and can degrade radio communications, causing radio blackouts and interference with GPS and satellite communications. Electrons and protons are accelerated towards the interplanetary space and can release their energy in the satellite electronics producing malfunctions and also the loss of the satellite itself.
SFs occurs typically in association with Coronal Mass Ejection (CME) and Solar Energetic Particle (SEPs)  events on the ground \cite{Papaioannou2016}. They occur at release of the CME from the Sun and contribute as a trigger mechanisms of their release.

During a SF, the magnetic reconnection originates a huge release of energy at the top of a magnetic loop. Particles are accelerated along the magnetic field lines towards the lower layers of the solar atmosphere and the interplanetary space.
The energy spectrum of a SF is dominated by three components below 100 keV:
\begin{itemize}
\item emission lines below 10 keV;
\item thermal Bremsstrahlung (expected to be weakly polarised)\cite{Emslie1980a};
\item non-thermal Bremsstrahlung emerging from about 10--30 keV\cite{Zharkova2010}
\end{itemize}
Theoretical models predict high linear polarisation of the non-thermal component, depending on the particle beaming and magnetic field properties \cite{Zharkova2010,Jeffrey2020}. Polarisation allows to disentangle different beaming models that could be degenerate if assessed only through spectroscopy.  Moreover, the directional properties of accelerated particles, relevant for Space Weather, can be derived from polarisation measurements\cite{Zharkova2010}.
CUSP is aimed to assess the linear polarization of the non-thermal component in the 25--100 keV energy range.

CUSP results are intended to contribute to the present and future networks for Space Weather, including the future ASI SPace weather InfraStructure (ASPIS) \cite{Plainaki2018}. SF photons reach the Earth few minutes after the flaring event, anticipating the CME particles by some hours up to some days. Polarimetry  rapidly gives information about the initial conditions of the detaching CME that are useful to  define the initial condition of the CME onset.

CUSP is currently approved for a Phase B study by the Italian Space Agency in the framework of the Alcor program aimed to develop CubeSat technologies and missions.
INAF-IAPS is the Prime Contractor and Principal Investigator of the scientific payload (with contributions from INAF-OAS Bologna and INAF-OAR). The payload front-end and back-end electronics is going to be designed and realized by SCAI Connect s.r.l. 
The 6U CubeSat platform is going to be designed and produced by IMT s.r.l. The Interdepartmental Center for Aerospace Industrial Research (CIRI-AERO) of the University of Bologna is performing the mission analysis, while the University of Viterbo ``La Tuscia" will take care of the ground segment, with a ground station located in the University campus.

\section{The payload: The hard X-ray polarimeter}
\label{sec:payload}
\begin{figure} [ht]
\begin{center}
\begin{tabular}{c}
\includegraphics[height=10cm]{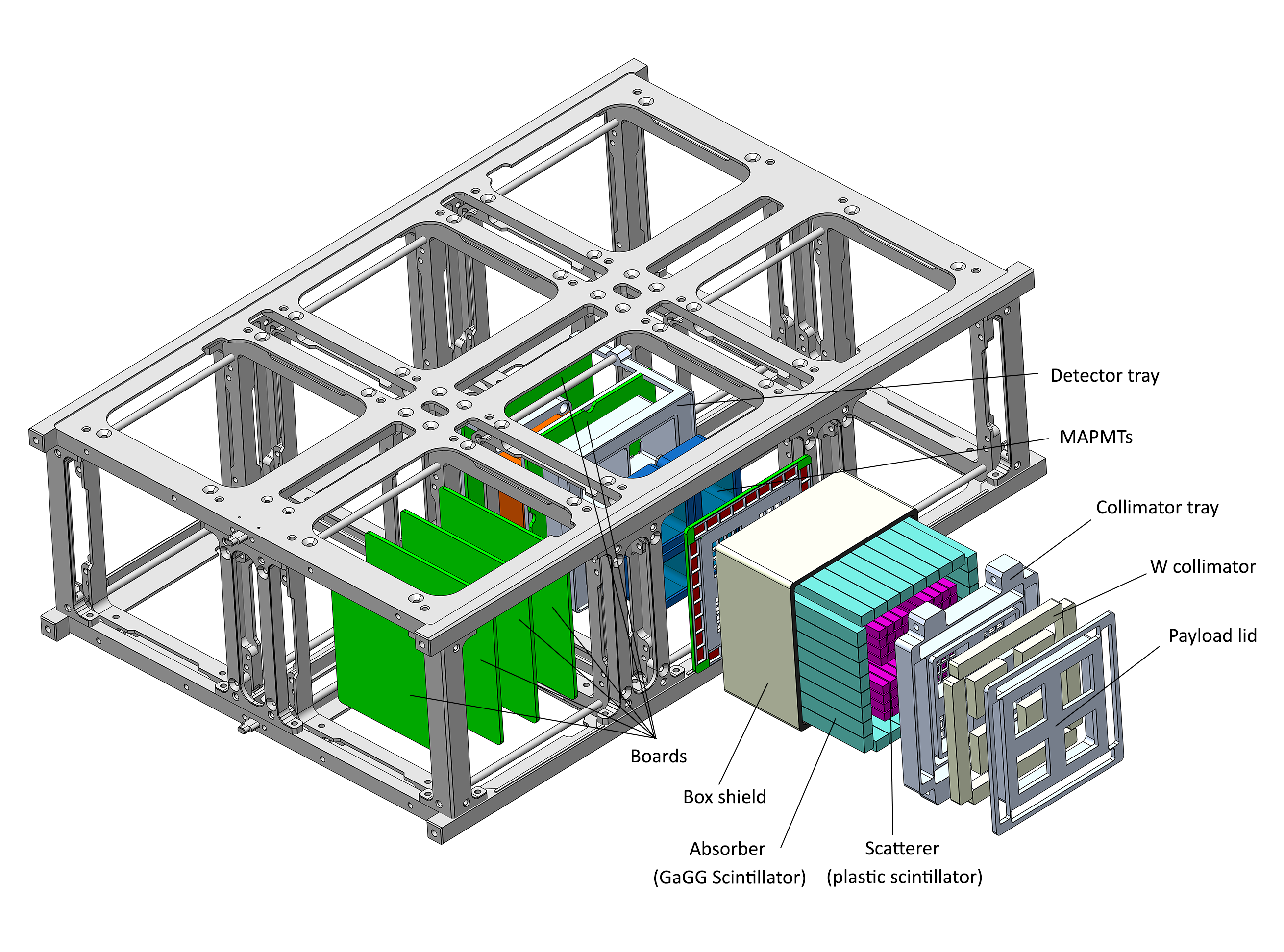}
\end{tabular}
\end{center}
\caption[example]
{CUSP payload as designed at the end of the Phase A represented in the CubeSat structure. \label{fig:views}}
\end{figure}

The CUSP payload comprises a dual-phase Compton scattering polarimeter (operating in the 25--100 keV energy band). The polarimetric measurement is performed by recording coincident events between plastic and inorganic scintillator rods (see Fig.~\ref{fig:views}). The 64 plastic scintillator rods allow to maximise the scattering probability (thanks to their low atomic number) with respect to the heavier 32 inorganic crystals made of GAGG:Ce (Gd$_3$Al$_2$Ga$_3$O$_{12}$:Ce) that maximise the photoelectric absorption of the scattered photon. The azimuthal angular distribution of plastic/GAGG coincidences is employed to perform polarimetry. Polarised radiation induces a preferential azimuthal angular direction of scattering (normal to the incident beam axis) as described by the Klein-Nishina cross section \cite{Heitler1954}:
\begin{equation}
\frac{d\sigma}{d\Omega}=\frac{{r_0}^2}{2}\frac{{E^\prime}^2}{{E}^2}\Biggr[ \frac{E}{E^\prime}+\frac{E^\prime}{E}-2\sin^2 \theta \cos^2 \phi \Biggl] \label{eq:KN}
\end{equation}
where
\begin{equation}
\frac{E'}{E}=\frac{1}{1+\frac{E}{m_e c^2}(1-\cos \theta)}\label{eq:EsuE}
\end{equation}
$E$ and $E^\prime$ are the energies of the incident and scattered photons, respectively. The polar scattering angle is $\theta$, measured from the incident photon direction. The angle $\phi$ is the azimuthal one measured from the plane identified by the incoming direction and the electric vector of the incident photon. Linearly polarised photons are preferentially scattered perpendicularly to their polarisation direction producing a modulation in the histogram of the $\phi$ angle.
The higher the modulated response, the higher the sensitivity of the polarimeter for a given polarization degree.
Such a sensitivity is defined by the modulation factor $\mu(\theta)$ (fraction of modulated signal corresponding to 100$\%$ polarised radiation):
\begin{equation}
\mu(\theta)=\frac{N_\mathrm{max}(\theta)-N_\mathrm{min}(\theta)}{N_\mathrm{max}(\theta)+N_\mathrm{min}(\theta)}=\frac{(\frac{d\sigma}{d\Omega})_{\phi=\frac{\pi}{2}}-(\frac{d\sigma}{d\Omega})_{\phi =0}}{(\frac{d\sigma}{d\Omega})_{\phi=\frac{\pi}{2}}+(\frac{d\sigma}{d\Omega})_{\phi =0}}=\frac{\sin^2\theta }{\frac{E}{E^\prime}+\frac{E^\prime}{E}-\sin^2 \theta} \label{eq:Muphi}
\end{equation}
For an energy of the incident photon $E<<m_ec^2=511$~keV (coherent scattering) $E = E^\prime$ and the maximum modulation factor is obtained for $\theta=90^\circ$ (orthogonally to the incident photons direction). By increasing the energy it occurs at narrower scattering angles (forward folding). However, at 100~keV it is still $\theta \simeq 90^\circ$ \cite{Fabiani2012c}.

The payload of CUSP comprises a tungsten collimator for limiting the field of view around the solar direction, the plastic and the inorganic scintillators assemblies, the light signals of which is readout by means of 4 Multi-Anodes Photomultiplier Tubes (MAPMTs) for a total of 64 channels (to readout plastic rods) and 32 Avalanche Photo-Diodes (APDs, to readout GAGG rods), respectively. Readout sensors have been selected to have a high heritage due to the very short implementation time required for the project. We selected APDs from past TSUBAME mission, unfortuantely lost in 2015,  (and similar MAPMTs) by Hamamatsu \cite{Yatsu2014}.

\begin{figure} [ht]
\begin{center}
\begin{tabular}{c}
\includegraphics[height=8cm]{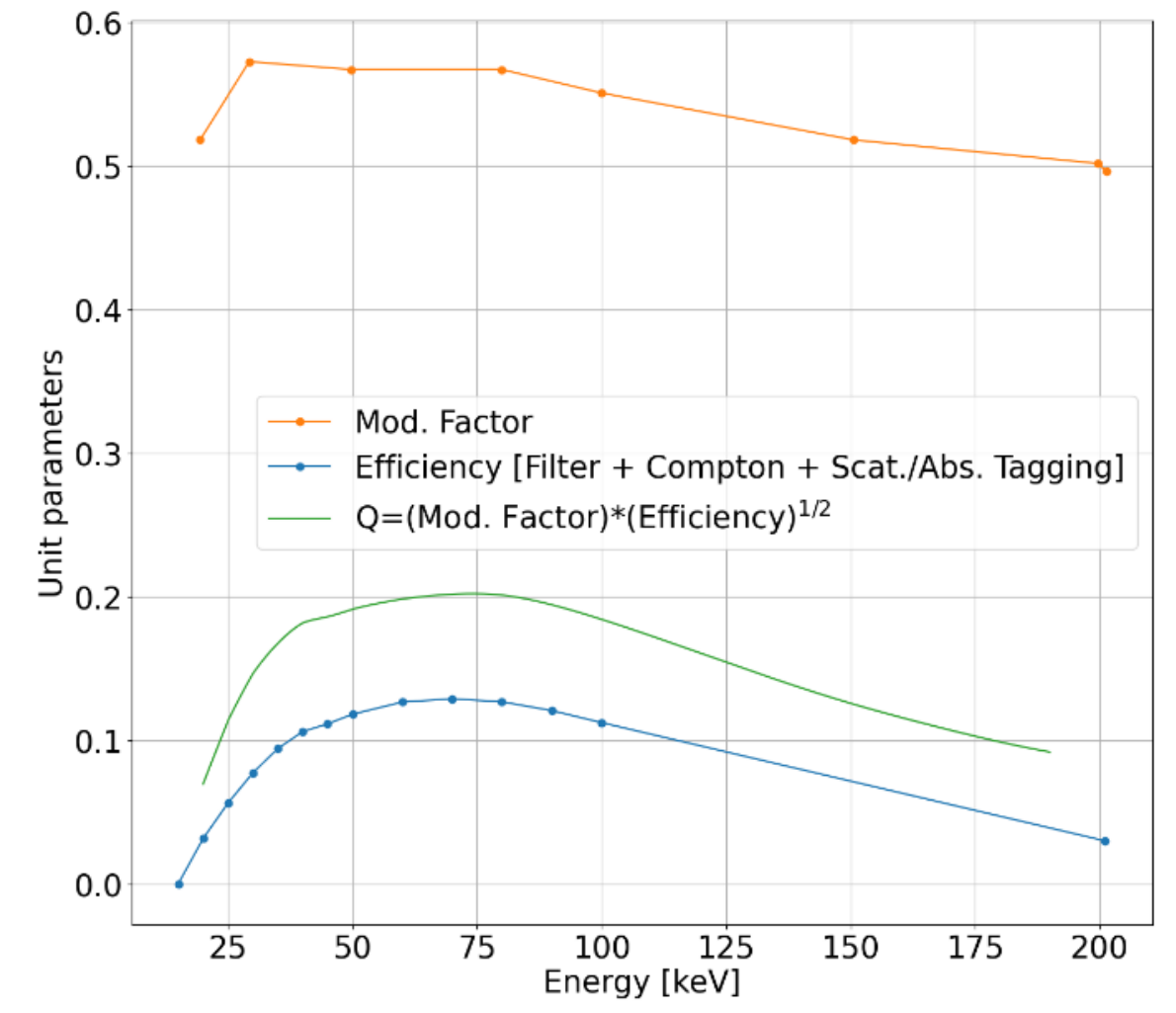}
\end{tabular}
\end{center}
\caption[example]
{Current best estimates of the modulation factor, efficiency (Compton interaction and tagging efficiency) and quality factor. \label{fig:curves}}
\end{figure}

The current best estimates of the modulation factor $\mu$ (modulation for 100$\%$ polarised radiation), efficiency $\epsilon$ (Compton interaction and tagging efficiency) and quality factor $Q$ are reported in Fig.~\ref{fig:curves}. Tagging efficiency is defined as the probability to detect an event in the scatterer after a detection of an event in the absorber \cite{Fabiani2012c}. The quality factor is defined as:
\begin{equation}
Q=\mu \sqrt{\epsilon}\label{eq:Q}
\end{equation}
It allows to identify the energy range in which the polarimeter is effective in measuring polarisation.
This parameter is derived from the Minimum Detectable Polarisation (MDP)\cite{Weisskopf2010} by assuming a source dominated observation (if background is negligible).

Table~\ref{tab:mdp} reports the Current best estimate (end of Phase A) of the MDP in the 25--100~keV energy band based on benchmark solar flares from Saint-Hilaire et al. (2008)\cite{SaintHilaire2008}.
Few minutes of integration time allow to measure the polarisation of solar flares that is expected to be well above the MDP (at a level of some tens of per cent).
\begin{table}[ht]
\caption{Current best estimate (end of Phase A) of the MDP in the 25--100~keV energy band based on benchmark solar flares from Saint-Hilaire et al. (2008)\cite{SaintHilaire2008}}
\label{tab:mdp}
\begin{center}      
\begin{tabular}{|c|c|c|c|}
\hline
\rule[-1ex]{0pt}{3.5ex} Flare Class & Integration Time (s) & Rate (cts/s) &MDP ($\%$)  \\
\hline
\hline
\rule[-1ex]{0pt}{3.5ex} M 5.2 & 284 & 20.4 &10.2  \\
\hline
\rule[-1ex]{0pt}{3.5ex} X 1.2 & 240 & 95.3 &5.0  \\
\hline
\rule[-1ex]{0pt}{3.5ex} X 10 & 351 & 1289.2 &1.1  \\
\hline
\end{tabular}
\end{center}
\end{table}

\section{The mission concept}
\label{sec:concept}

\begin{figure} [ht]
\begin{center}
\begin{tabular}{c}
\includegraphics[height=7cm]{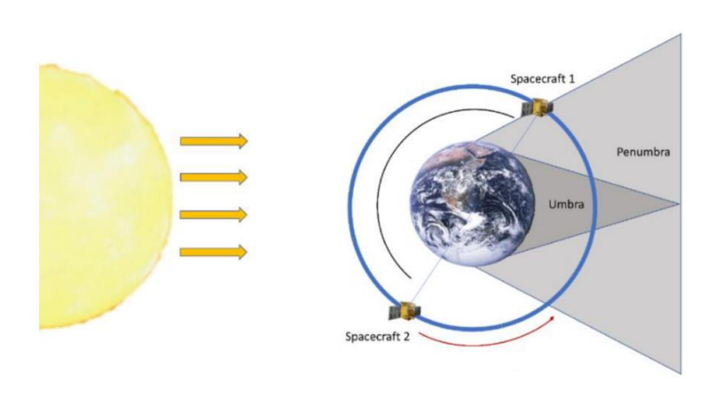}
\end{tabular}
\end{center}
\caption[example]
{CUSP constellation when CubeSats are at the maximum phase angle. \label{fig:orbit}}
\end{figure}

The CUSP mission foresees a constellation consisting 2 CubeSats orbiting the Earth in the same plane with a time variable phase difference such to allow a significant observation time of the Sun (average during the Nominal Operation Phase $> 44\%$, if at least one satellite is launched, see Fig.~\ref{fig:orbit}). For a Mid-Morning SSO orbit at 600~km height the average fraction of time of Sun observability during the Mission Operation Phase is about $\sim 70\%$, whereas for a Dawn-Dusk orbit at the same altitude this time fraction increases to $88 \%$.
This configuration allows to have one or two satellites available to observe the Sun, except when both are prevented to perform the observation due to the passages in the South Atlantic Anomaly (SAA) or in the Polar Exclusion Belts (PEB north and south), or during contacts with the ground station and eclipses.
Moreover, for some fraction of time that depends on the specific orbit and phase difference between two satellites, they can observe the Sun at the same time. In this case the sensitivity increases by a factor $\sqrt{2}\sim 40\%$, since the effective area available is doubled.
During Phase B a trade-off analysis will be performed to allow ASI to decide if to proceed towards the launch with one or two CubeSats.

During the Sun observation each satellite rotates around the Sunward direction of pointing. The rotation allows to reduce the systematic effects known as spurious modulation. The duration of a SF peak of intensity in the hard X-rays lasts few minutes. A rotation speed of at least 1~RPM is sufficient for sampling the modulation curve induced by polarization on a time scale of 30 seconds (half of the rotation period).

During data downlink each satellite will interrupt the observation and will transition to a in 3-axis stabilized configuration for optimizing the transmission link.

\section{The platform}
\label{sec:platform}
\begin{figure} [ht]
\begin{center}
\begin{tabular}{c}
\includegraphics[height=7cm]{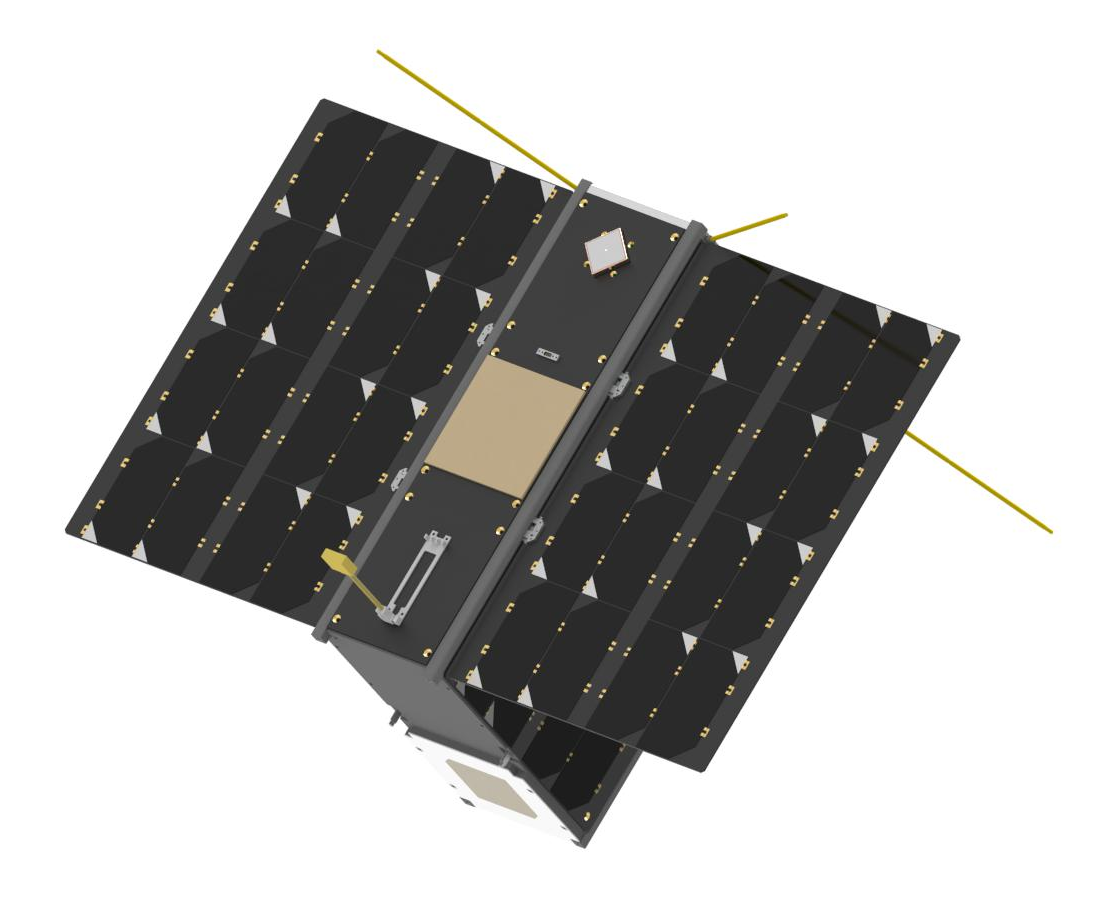}
\end{tabular}
\end{center}
\caption{CUSP satellite.\label{fig:platform}}
\end{figure}

Each CUSP satellite (see Fig.~\ref{fig:platform}) is based on a CubeSAT 6U platform. The architecture is developed by IMT s.r.l., on the basis of its consolidated experience in the realization of different CubeSats, also in the institutional framework. The 6U CubeSat is based on the heritage of the HORTA and EOSS platforms (6U CubeSat platforms funded by Italian regional POR / FESR 2014-20 projects of Lazio and Puglia regions, respectively).
The telemetry and remote controls are realized through UHF band uplink and downlink communication with omnidirectional turnstile antennas. The S-band communication subsystem, in support of the UHF one, provides a higher downlink speed to allow scientific data transfer.
The preliminary performance of the CUSP platform are reported in Table~\ref{tab:platform}.

\begin{table}[ht]
\caption{Preliminary performance of the CUSP platform.}
\label{tab:platform}
\begin{center}      
\begin{tabular}{|c|c|}
\hline
\rule[-1ex]{0pt}{3.5ex} Peak Power & $\sim 30$~W with Deployable Panels in Sun Pointing\\
\hline
\rule[-1ex]{0pt}{3.5ex} Battery & Up to 84 Wh (baseline 42 Wh) \\
\hline
\rule[-1ex]{0pt}{3.5ex} Pointing accuracy & $\pm2^\circ$~\emph{@}~$1\sigma$  \\
\hline
\rule[-1ex]{0pt}{3.5ex} Operative frequencies & S-Band downlink;
UHF-Band uplink / downlink \\
\hline
\rule[-1ex]{0pt}{3.5ex} Downlink  throughput & Up to 5 Mbps   \\
\hline
\rule[-1ex]{0pt}{3.5ex} Available interfaces & CAN Bus, I2C, UART, SPI, RS485  \\
\hline
\rule[-1ex]{0pt}{3.5ex} Regulated bus & 3,3V, 5V e 12V   \\
\hline
\rule[-1ex]{0pt}{3.5ex} Not regulated bus & 16V (12V-16.8V)   \\
\hline
\rule[-1ex]{0pt}{3.5ex} Available volume for the payload & 2.5U   \\
\hline
\rule[-1ex]{0pt}{3.5ex} Nominal life time & 3 years in LEO   \\
\hline
\end{tabular}
\end{center}
\end{table}

\section{The ground station}
\label{sec:ground}

\begin{figure} [ht]
\begin{center}
\begin{tabular}{c}
\includegraphics[height=5cm]{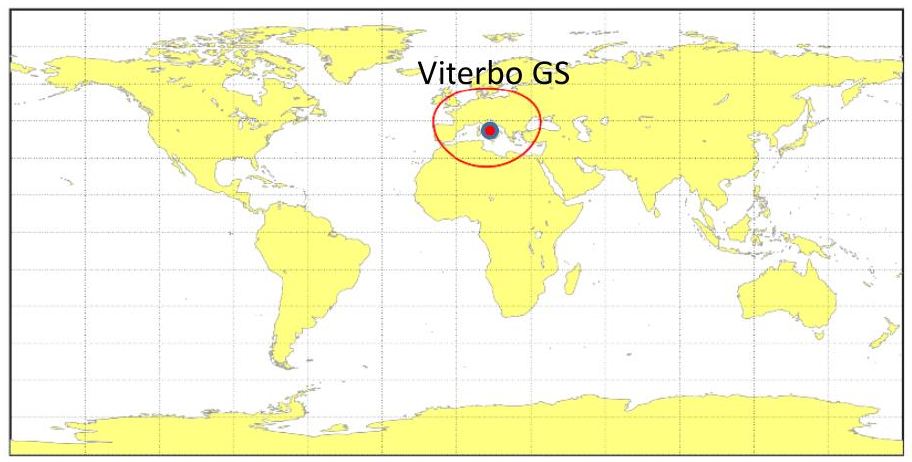}
\end{tabular}
\end{center}
\caption{The ground station at the ''La Tuscia" University of Viterbo. \label{fig:groundstation}}
\end{figure}

The Ground Station (see Fig.~\ref{fig:groundstation}) of the CUSP mission is located at the “La Tuscia” University of Viterbo, on the building F of the ``Riello" Campus.
It was built in 2019 as part of the HORTA project (Italian regional funds POR-FESR 2014-2020 of Lazio region). The Station allows for autonomous satellite tracking (using TLE satellite data - Two Lines Elements) and  satellite communication.

The available antennas and bands are:
\begin{itemize}
\item VHF: Uplink and Downlink
\item UHF: Uplink and Downlink
\item S-band: Downlink
\end{itemize}
The UHF/VHF bandwidth are 9.6 kbps as default for downlink (available also 1.2/ 2.4 / 4.8 kbps)
and 1.2 kbps as default for uplink (available also 2.4 / 4.8/ 9.6 kbps).
The S-band bandwidth is up to 1 Mbps for downlink.
The pointing accuracy of the ground station is 0.1° (both azimuth and elevation) with a minimum tracking speed of 2$^\circ$/s in azimuth, 1.8$^\circ$/s in elevation.
Data received from the ground station is transferred via fibre optics cable to dedicated workstations in the Mission Control Center that allows to schedule the passage of the satellite, the
TT\&C and Payload data transceiving activities. Moreover, it provides a Network Server service for data delivery to third parties.
From the preliminary Mission Analysis, about 3--4 contacts per day will be possible during mission operations.

\section{The planning}
\label{sec:planning}

The CUSP mission is based on high TRL subsystems. The Platform and the Ground Station can exploit a large heritage that allow them to guarantee a TRL 7. The Payload  is based on elements with a high TRL (MAPMT, APD, scintillators, ASICs, coincidence technique), but the polarimetric detector as a whole needs to be implemented. Thus, a TRL 3 is quoted.
The Model Philosophy is based on the production of one detector prototype and a mechanical model at the end of Phase B. The detector prototype will be representative of the front-end functionalities to enhance detector TRL from 3 to 4.
Then one payload EQM (engineering qualification model) will be designed during phase B to be produced and tested during phase C. It will be representative of the payload to enhance detector TRL from 4 to 7.
From the satellite point of view, 2 CubeSats will be produced:
\begin{itemize}
\item 1 Proto-flight Model (PFM). To be qualified at proto-qualification level
\item 1 Flight Model (FM). To be qualified at acceptance level.
\end{itemize}
The Calibration of the Hard X-ray Polarimeter of each CubeSat will be carried out at INAF-IAPS calibration facility (already employed for calibrating the IXPE Detector Units)\cite{Muleri2021ice} and possibly to Synchrotron test beamlines. The INAF-IAPS calibration facility is going to be extended up to hard X-rays by employing X-ray tubes up to 100-150 KV for direct beams and scattered ones (polarized radiation)
A 12 months phase B is planned to start in mid-2024.

\section{Conclusion}

Solar flares can represent a threat for human activities in space and on ground. They are usually associated to Solar Energetic Particles Events (SEPs) at the Earth and Coronal Mass Ejections (CMEs) that originate geomagnetic storms. X-ray polarimetry would allow to assess magnetic reconnection on the Sun and beaming properties during particle acceleration along the solar magnetic field lines. CUSP mission is aimed to measure the linear polarisation of solar flares in the 25-100 keV energy band to probe such physical processes. CUSP will allow to measure X-ray polarization of an X1-class Solar Flare with an MDP of $5\%$ contributing to the understanding of these solar violent phenomena. It will also participate in the present and future networks for Space Weather, including the ASI SPace weather InfraStructure (ASPIS).

\acknowledgments 
Activity funded by ASI phase A contract 2022-4-R.0.


\bibliographystyle{spiebib} 

\end{document}